# Charge-transfer-mediated boron magneto-ionics: Towards voltage-driven multi-ion transport

Z. Ma[1,*], K.-A. Kantre[2], H. Tan[1], M. O. Liedke[3], J. Herrero-Martín[4], E. Hirschmann[3], A. Wagner[3], A. Quintana[1], E. Pellicer[1], J. Nogués[5,6], J. Meersschaut[2], J. Sort[1,6,*], E. Menéndez[1,*]

[1]Departament de Física, Universitat Autònoma de Barcelona, E-08193 Cerdanyola del Vallès, Spain. [2]IMEC, Kapeldreef 75, B-3001 Leuven, Belgium. [3]Institute of Radiation Physics, Helmholtz-Zentrum Dresden – Rossendorf, Dresden 01328, Germany. [4]ALBA Synchrotron Light Source, 08290 Cerdanyola del Vallès, Spain. [5]Catalan Institute of Nanoscience and Nanotechnology (ICN2), CSIC and BIST, 08193 Barcelona, Spain. [6]Institució Catalana de Recerca i Estudis Avançats (ICREA), Pg. Lluís Companys 23, E-08010 Barcelona, Spain.

*e-mail: ma.zheng@uab.cat (Z. Ma), jordi.sort@uab.cat (J. Sort), enric.menendez@uab.cat (E. Menéndez)

Voltage control of magnetism via magneto-ionics —where ion transport and/or redox processes drive magnetic modulation— holds great promise for next-generation memories and computing. This stems from its non-volatility and ability to precisely tune both the magnitude and speed of magnetic properties in an energy-efficient manner. However, expanding magneto-ionics to incorporate novel mobile ions or even multiple ion species is crucial for unlocking new phenomena and enabling multifunctional capabilities. Here, we demonstrate voltage-driven multi-ion transport in a FeBO system with increasing oxygen content, progressively transitioning from an electrostatic-like response to a more pronounced electrochemical (magneto-ionic) behavior. The voltage-driven transport of both B and Fe is activated by oxidation state tuning, owing to the larger electronegativity of oxygen. Such charge-transfer effects allow multi-ion magneto-ionics, where O ions move oppositely to Fe and B ions. These results pave the way for programmable functionalities by leveraging elements with different electron affinities through charge-transfer engineering.

The use of voltage to control magnetism through converse magneto-electric effects has emerged as a promising approach for developing energy-efficient spintronic devices. Among the different magneto-electric mechanisms, magneto-ionics —where voltage-driven ion transport and/or redox processes modulate magnetism— stands out due to its non-volatility and versatility in modifying the magnitude and speed of magnetic response[1]. Thus far, magneto-ionic systems/devices have primarily relied on single-ion manipulation, such as $H^+$[2], $Li^+$[3], $O^{2-}$[4,5], $F^-$[6], $OH^-$[7], or $N^{3-}$[8]. Even though dual-type magneto-ionics has been demonstrated[9-11], expanding the range of mobile ions and enabling multi-ion transport are pivotal for achieving integrated multifunctionality and on-demand programmable magneto-ionics.

Here, we report on the foremost demonstration of voltage-driven movement of B ions in FeBO-based heterostructures, activated by engineering the oxidation state of B through oxygen incorporation. The presence of oxygen has a synergetic two-fold effect: (i) it alters the oxidation state of B in FeBO heterostructures, shifting it toward a more positive value as the oxygen-to-argon pressure ratio increases during sputtering —an effect driven by oxygen's higher electronegativity, and (ii) it increases resistivity, enhancing electric field penetration and facilitating voltage-driven ion transport[12]. In FeB, ion diffusion is mainly restricted to the top layers due to surface passivation. The system is weakly magneto-ionic, showing an electrostatic-like response[13]. However, as oxygen content increases in FeBO heterostructures, the electrochemical (magneto-ionic) behavior reinforces because of the increased resistivity, where oxygen ions migrate in the opposite direction to Fe and B ions. Namely, an unprecedented voltage-induced multi-ion magneto-ionic mechanism is observed, which not only alters the magnetization but also increases coercivity upon voltage actuation.

50 nm-thick FeBO films were sputtered at room temperature on Si(100) wafers, previously coated with 15 nm Ti, 50 nm Au, and 10 nm Ta, under varying $O_2$/Ar gas flow ratios (*i.e.*, $\frac{O_2\ flow}{O_2\ flow\ +\ Ar\ flow}$) — 0 (no oxygen), 2, and 5% $O_2$, denoted hereafter as FeB, FeBO (2% $O_2$), and FeBO (5% $O_2$), respectively (Methods). The microstructure of the FeB and FeBO (2% $O_2$) is quite different. FeB presents an amorphous-like homogeneous morphology (Fig. S1), while FeBO (2% $O_2$) shows a columnar-shaped structure with elongated grains of about 3 nm in diameter (Fig. 1a), triggered by the presence of oxygen[14]. Moreover, the high-angle annular dark-field scanning transmission electron microscopy (HAADF-STEM) and elemental electron energy loss spectroscopy (EELS) mappings of the FeBO (2% $O_2$) sample (Fig. 1b) show that, apart from a top overoxidized 2 - 3 nm layer (due to the exposure to air), the Fe and O are homogenously distributed throughout the film. The B signal, which is challenging to measure using EELS, is at the background level and is thus not shown. The fast Fourier transform (FFT) pattern in Fig. 1c shows predominantly a diffuse diffraction ring and a hallo, evidencing the highly nanostructured nature of the film, together with some diffraction spots, all compatible with a phase close to metallic Fe (PDF® 00-001-1262). This is consistent with the lattice fringes detected in sporadic regions of the as-grown FeBO (2% $O_2$) sample (see Fig. S2).

To investigate magneto-electric effects of the different samples, room-temperature magnetometry measurements were carried out while electrolyte gating the heterostructures (see Fig. 1d and Methods). Fig. 1e displays the evolution of the change in saturation magnetization, $\Delta M_S$, while applying a gate voltage of −50 V, for the FeB and FeBO (2% $O_2$) heterostructures. Both films exhibit ferromagnetism in the as-grown state (Figs. 1f,g)[15], and their $\Delta M_S$ increase as a response to biasing. Subsequently, after gating for 1 h, the voltage was turned off and hysteresis loops were recorded. As shown in Fig. S3, the increase in saturation magnetization ($M_S$) is permanent (non-volatile) for the FeBO (2% $O_2$) sample, compatible with the partial reduction of FeBO to metallic FeB (*i.e.*, magneto-ionics). In contrast, the voltage-induced $M_S$ increase in the FeB sample is less pronounced and shows volatility, recovering the initial value of the as-grown sample during the duration of the hysteresis loop (Fig. S3), indicating that FeB exhibits minimal magneto-ionic behavior. As can be seen in Fig. 1f, the coercivity ($H_C$) of the electrically-treated FeB slightly decreases by 9% upon voltage actuation. This decrease in $H_C$ is a characteristic magneto-electric effect observed in electrostatic-responsive materials, such as metallic thin films like FePt[16], and porous Cu-Ni[17], as well as in systems that exhibit a combination of electrostatic and electrochemical behaviors[18]. In electrostatic systems, coercivity reduction is typically attributed to voltage-induced modifications in electronic density, whereas, in mixed systems, it is primarily driven by magneto-ionic effects, where ion transport and/or redox processes play a dominant role. When a voltage polarity favors the reduction of oxidized phases, the resulting decrease in coercivity is generally linked to an increase in the ferromagnetic fractions. This enhances percolation and strengthens exchange interactions among ferromagnetic regions, which ultimately leads to a decrease in coercivity[19]. Conversely, the coercivity of the electrically-treated FeBO (2% $O_2$) sample increases by 16% (Fig. 1g), opposing the trend of oxide-based systems with mixed responses. To serve as a control sample, a FeO (2% $O_2$) film without B was electrolyte-gated, resulting in a reduction in coercivity (Fig. S4). This suggests that B plays a crucial role in the unusual magneto-electric response of FeBO (2% $O_2$).

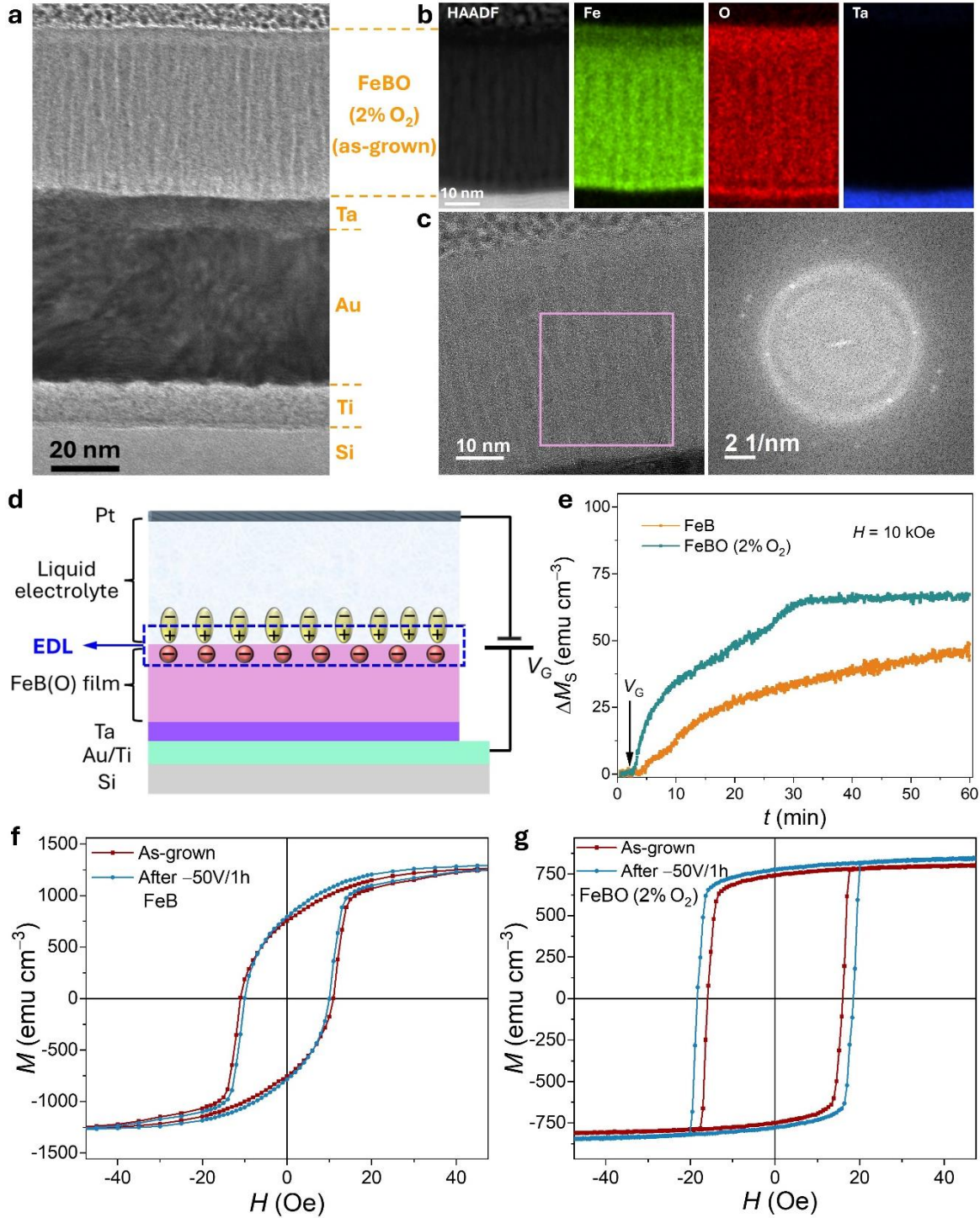

**Fig. 1 | Voltage control of magnetism in FeB- and FeBO-based heterostructures. a**, Cross-sectional TEM image of the as-grown FeBO (2% $O_2$) heterostructure, showing the columnar growth of the film. **b**, HAADF-STEM micrograph of the heterostructure and the corresponding Fe, O, and Ta elemental EELS mappings. **c**, High-resolution TEM image of the FeBO layer, together with the FFT pattern of the region enclosed by the red square. **d**, Schematic of the sample and electrolyte gating configuration used. As a consequence of biasing, an electric double layer (EDL) forms at the film/electrolyte interfaces. **e**, Variation of saturation magnetization (*i.e.*, $\Delta M_S$) as a function of time *t* while applying voltage. Note that the curves were recorded while an in-plane magnetic field of 10 kOe was applied. Gate voltages, $V_G$, were applied starting from *t* = 2 min. **f, g** Comparison of the hysteresis loops of the as-grown and gated (–50 V for 1 h) FeB and FeBO (2% $O_2$) samples, respectively.

To disentangle the role of B in the magneto-electric behavior of the heterostructures, time of flight-energy (TOF-E) elastic recoil detection analysis (ERD) was carried out to determine the composition across the samples. As seen in Figs. 2a,b, the concentration of Fe and B (in atomic percentage, at.%) as a function of depth is plotted for the as-grown and voltage-treated FeB and FeBO (2% $O_2$) samples, respectively. In the FeB sample, the Fe and B depth profiles of the as-grown and electrically-treated samples coincide across the thickness of the films. In contrast, the Fe and B profiles of the as-grown and voltage-treated FeBO (2% $O_2$) samples do not overlap. Instead, there is a clear shift of the voltage-treated profiles towards deeper regions of the sample, evidencing a voltage-induced transport of both the Fe and B ions. The transport of both Fe and B ions towards the negatively-biased Au electrode indicates their cationic nature. Unfortunately, TOF-E ERD cannot be used to determine the direction of voltage-driven oxygen transport, since it cannot distinguish between oxygen incorporated from the atmosphere[20] and oxygen that has undergone voltage-induced movement. However, strong evidence for the upward movement of the O ions under voltage actuation can be observed from the permanent increase in $M_S$, as shown in Fig. 1g and Fig. S3. This is ascribed to the voltage-driven transport of ions toward the liquid electrolyte, partly reducing the FeBO to FeB. Since the relative atomic fractions of B, with respect to the total amount of B and Fe, remain constant within the films (Fig. S5), the influence of stoichiometry on this permanent increase in $M_S$ can be ruled out. Additional evidence for the upwards movement of the O ions upon voltage actuation can be detected in the HAADF-STEM-EELS characterization of the voltage-treated sample, where it can be clearly observed that topmost oxide-rich layer (i) doubles in thickness (from 5 nm in the as-grown state —see Fig. 1b— to 10 nm after voltage treatment) and (ii) loses Fe after gating (Fig. 3).

To elucidate the role of the oxidation state in the evolution of magneto-electric properties, X-ray absorption spectroscopy (XAS) was carried out in fluorescence yield mode (FY; sensitive to the bulk of the sample). Figs. 2c,d show the B $K$ and Fe $L_{2,3}$ FY-XAS spectra, respectively. As seen in Fig. 2c, the peak located at around 191 eV of the B $K$-edge XAS spectra shifts towards higher energies with increasing oxygen content, indicating a higher oxidation state of B[21]. Since the B ion movement is observed in FeBO (2% $O_2$) but not in FeB, this suggests the existence of a threshold oxidation state required to induce B transport. This charge-transfer effect also takes place in Fe, manifested as the newly developed shoulders at the higher energy ends of the Fe $L_{2,3}$ main peaks (indicated by the downward arrows in Fig. 2d).

To further unravel the role of changes in the oxidation state in B, the magneto-electric properties of the FeBO (5% $O_2$) sample were also investigated. As shown in Fig. S6, FeBO (5% $O_2$) shows magneto-ionics, evidencing clear electrochemical effects, where the heterostructure evolves from paramagnetic to ferromagnetic upon gating. This is linked to the higher oxygen content and the concomitant higher resistivity, which enhances electric field penetration[12]. As shown in Fig. S7, for the as-grown FeBO (5% $O_2$) sample, the higher energy shoulder peaks observed in the FY-XAS spectra of the FeBO (2% $O_2$) sample become more intense, indicating an enhanced Fe oxidation state. Moreover, the peak in the B FY-XAS spectrum of the as-grown FeBO (5% $O_2$) exhibits a considerable shift towards higher energies, approaching the energy of boron oxide ($B_2O_3$), strongly suggesting the formation of $B_2O_3$ when the samples are grown in large amounts of oxygen. The high stability of $B_2O_3$[22] results in a system where B is less mobile compared to FeBO (2% $O_2$), as confirmed by the reduced shifts of B concentration towards deeper parts upon gating in FeBO (5% $O_2$) compared to FeBO (2% $O_2$) (see the ERD results in Fig. S8). Thus, the results indicate that, in the FeBO system, to induce multi-ion movement an optimal oxidation state of B is required. However, the optimization of the oxidation states of the different ions to induce multi-ion transport should depend on the ions involved and the material.

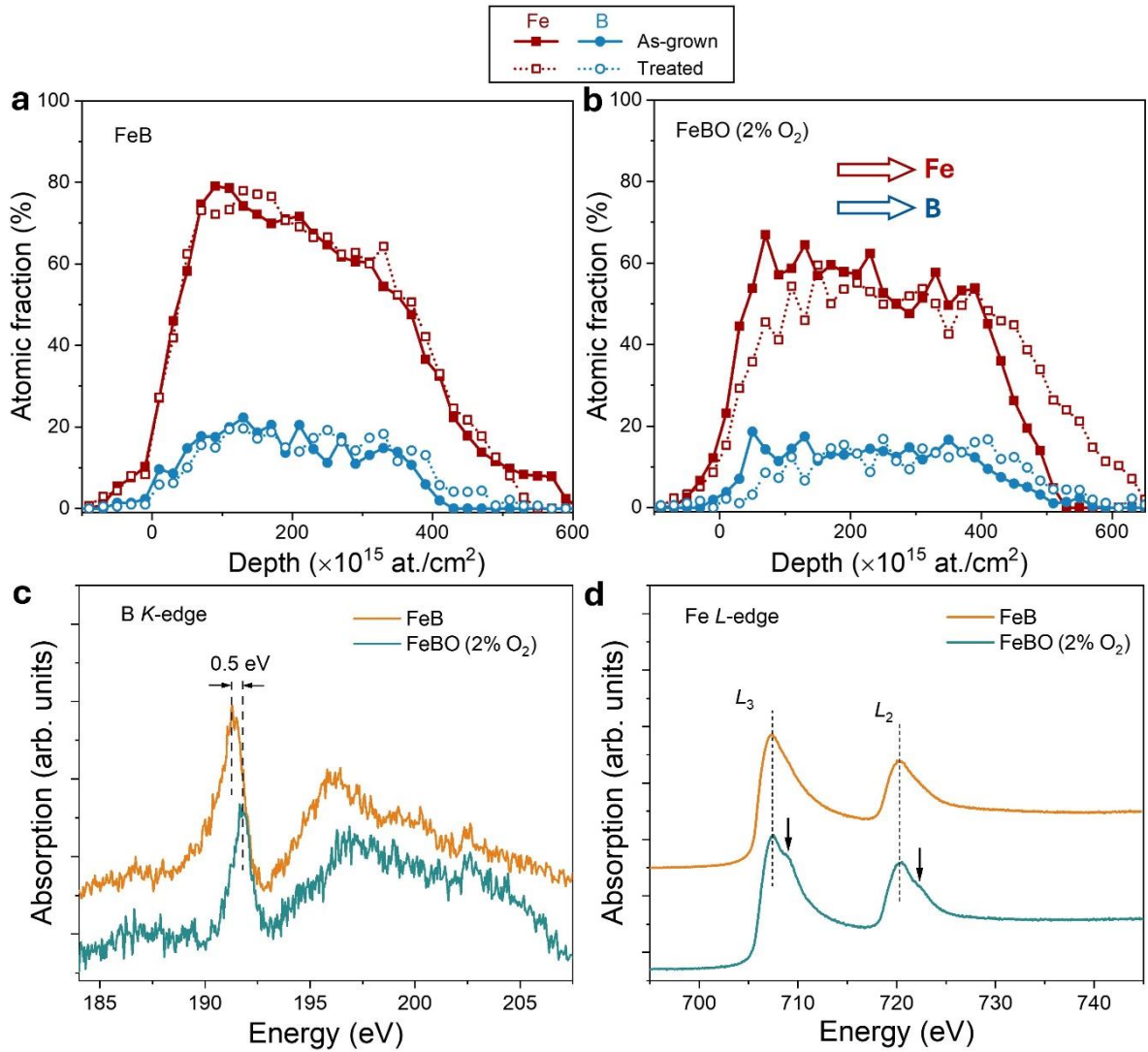

**Fig. 2 | Compositional characterization of the heterostructures by TOF-E ERD and XAS. a**, TOF-E ERD depth profiles of Fe and B elements for the as-grown and gated FeB films. **b**, TOF-E ERD depth profiles of Fe and B elements for the as-grown and gated FeBO (2% $O_2$) films. **c**, B $K$-edge, and **d**, Fe $L$-edge XAS spectra for the as-grown FeB and FeBO (2% $O_2$) films in fluorescence yield (FY) mode. The arrows in panel b indicate the direction of movement of Fe and B while gating.

A remarkable feature induced by the multi-ion movement in the FeBO (2% $O_2$) upon gating, is the observed increase in coercivity. To shed light on this effect, a detailed structural characterization was carried out by TEM and EELS (Fig. 3). Fig. 3a shows the cross-sectional TEM image of a FeBO (2% $O_2$) heterostructure after gating at −50 V for 1 h. The HAADF-STEM image and Fe, O, and Ta EELS maps of the cross-section confirm that the topmost layer is O-rich but Fe-depleted (Fig. 3b), in agreement with the voltage-driven transport of Fe ions downward toward the bottom electrode. Furthermore, the high-resolution TEM image and FFT analysis in Fig. 3c reveal that voltage application leads to an increased degree of nanostructuring of the films, with an amorphous-like character. This is evidenced by the disappearance of diffraction spots, leaving only a diffuse diffraction ring. A closer examination of the high-resolution images in Figs. 3d,e,f,g, which compare the top and bottom regions of the microstructures along with their corresponding FFT patterns, confirms that this voltage-induced structural modification occurs consistently throughout the films.

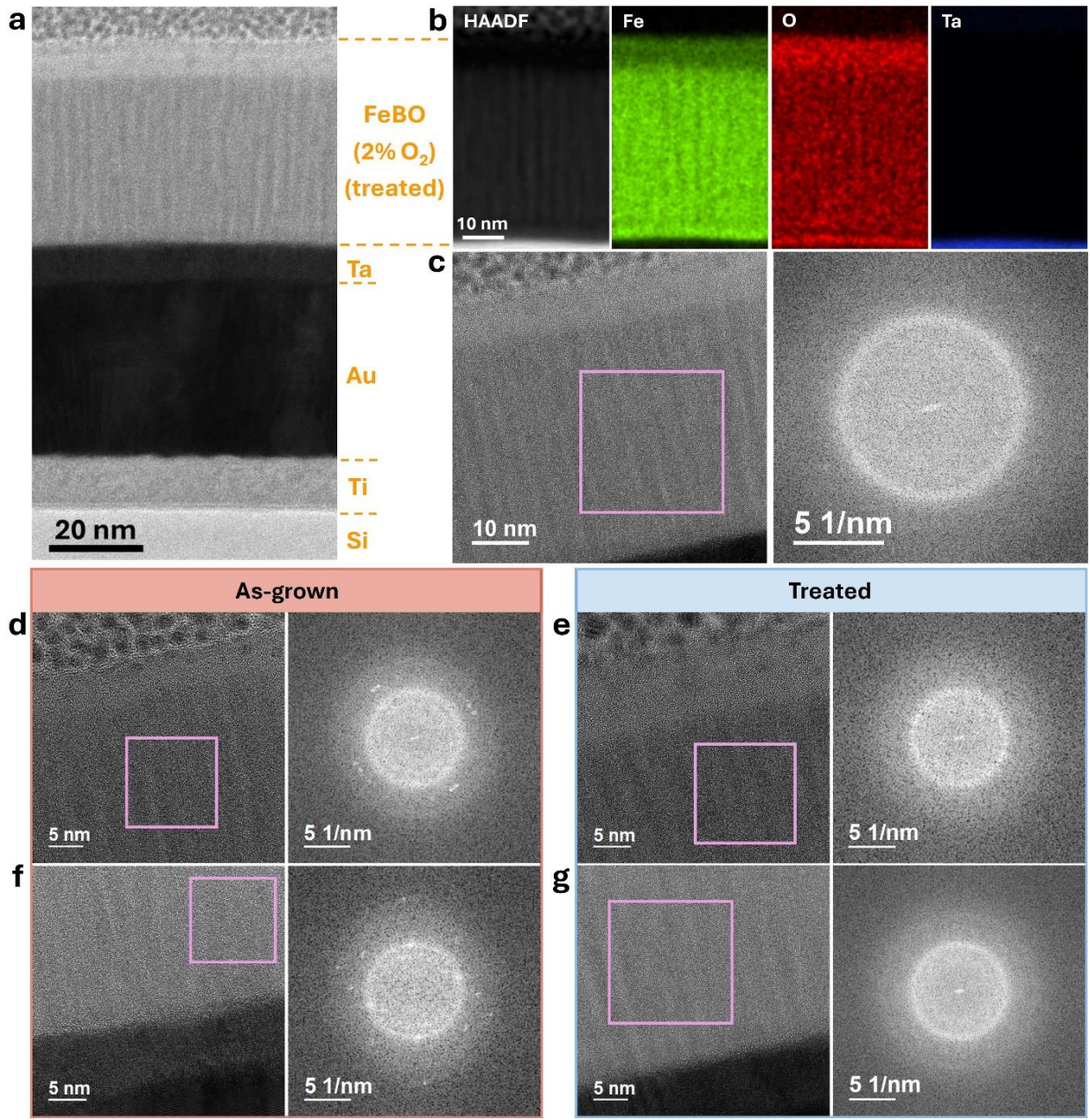

**Fig. 3 | Microstructural and compositional characterizations of the as-grown and voltage-actuated FeBO (2% O₂) films. a**, Cross-sectional TEM image of the FeBO (2% O₂) heterostructure after a gating actuation at −50 V for 1 h. **b**, Representative HAADF-STEM micrograph of the heterostructure and the corresponding Fe, O, and Ta elemental EELS mappings. **c**, High-resolution TEM image of the gated FeB-O layer, together with the FFT pattern of the region enclosed by the red square. **d-g**, Comparison of enlarged high-resolution TEM images of the top (**d**, **e**) and bottom (**f**, **g**) regions of the heterostructures in their

pristine (**d**, **f**) and gated (**e**, **g**) states. The corresponding FFT patterns for the regions marked with red squares are displayed in the right panels.

The coercivity of nanostructured systems typically decreases with reduced crystallinity (*i.e.*, finer grain size)[23]. However, in amorphous-like systems, an increase in inhomogeneous local atomic environments[24] and strains[25] can enhance magnetic anisotropy gradients, leading to increased coercivities. To better understand the origin of this coercivity increase, structural characterization at the atomic level was performed by variable energy positron annihilation lifetime spectroscopy (VEPALS, Fig. 4 and Supplementary Note 1). In FeBO (2% $O_2$) heterostructures, small vacancy-clusters (*i.e.*, $\tau_1$) are the majority defect type, with a relative intensity (*i.e.*, $I_1$) exceeding 55% (Fig. 4a,b). The size of vacancy agglomerations is larger in the sub-surface region, decreasing towards the buffer layer, which is likely related to the oxygen overstoichiometry and Fe depletion (indication of iron vacancies). The second lifetime component $\tau_2$ can be related with open volumes at the columnar grain boundaries and their intersections since its counterpart is smaller compared to $\tau_1$. Interestingly, the defect structure in the buried regions of FeBO (2% $O_2$) is altered by the biasing. Specifically, $I_1$ increases by more than 5% after voltage actuation, indicating a raise in the density of small vacancy clusters, which promotes the development of inhomogeneous local atomic environments and strains, enhancing magnetic anisotropy gradients, and ultimately leading to the observed increase in coercivity[24,25]. At the same time, a slight thickening of the defected sub-surface region is found in the relative intensity depth profile, in agreement with the TEM investigations (Fig. 3a,b).

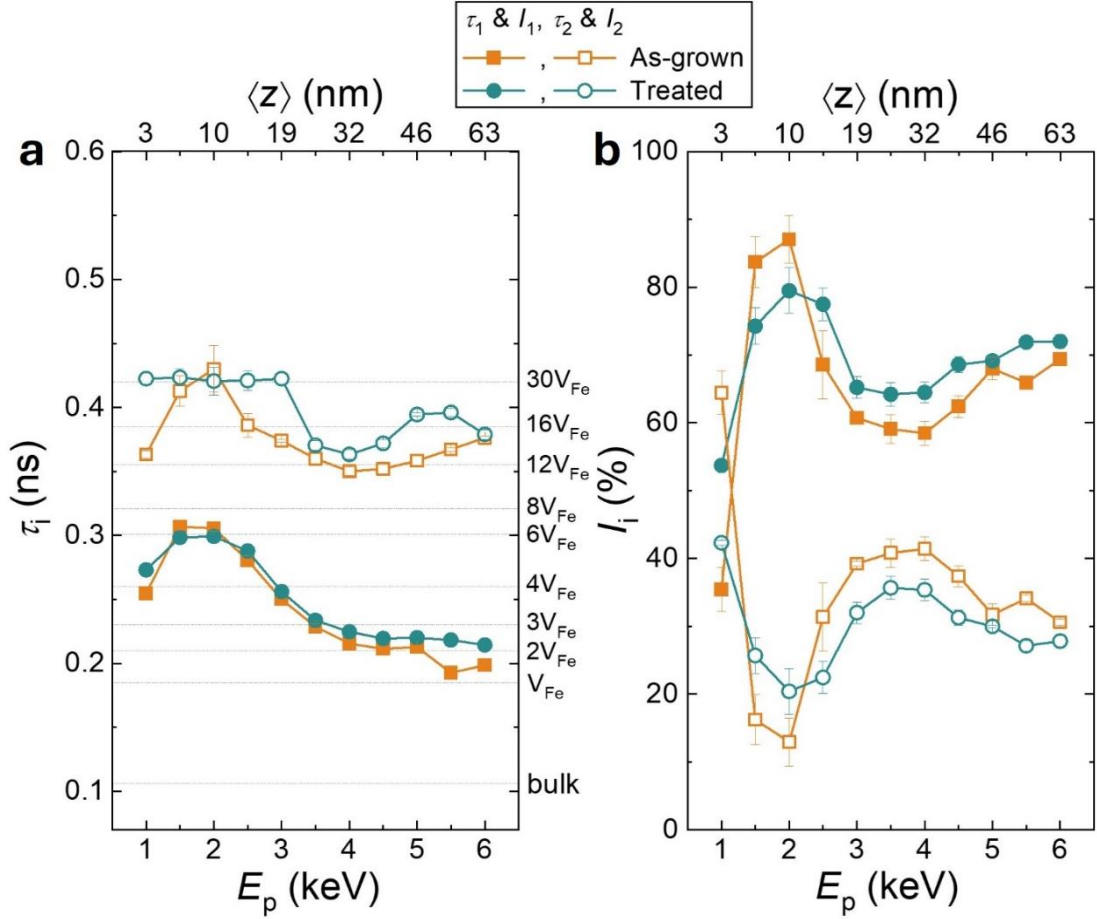

**Fig. 4 | Defect characterization of the FeBO (2% $O_2$) films in both pristine and voltage-actuated states by VEPALS. a**, Positron lifetime components ($\tau_1$ and $\tau_2$), and **b**, the corresponding relative intensities ($I_1$ and $I_2$), as a function of positron mean implantation depth ($\langle z \rangle$) and implantation energy ($E_p$) (shown in the top and bottom axes, respectively) for the as-grown and voltage-treated FeBO (2% $O_2$) films. See Supplementary Note 1 for further details on the correlation of positron lifetime with the number of Fe vacancies ($V_{Fe}$).

Our findings reveal a novel voltage-driven multi-ion (B, Fe, and O) transport in a FeBO system. As O content increases, the system transitions from an electrostatic-like (weakly electrochemical) response to a more pronounced electrochemical (magneto-ionic) behavior, driven by an increase in electrical resistivity. The voltage-driven transport of B, demonstrated for the first time, is enabled and triggered by tuning its oxidation state through oxygen incorporation. B magneto-ionics becomes activated within a specific oxygen concentration range. Lower oxygen levels result in weakly magneto-ionic systems with low resistivity, while higher oxygen concentrations lead to the formation of stable boron oxide, limiting voltage-driven B transport. In addition, the voltage-induced multi-ion movement leads to a remarkable coercivity increase, which is related to the increased degree of amorphization in voltage-treated samples. These findings provide a pathway for designing multi-ion manipulation through charge-transfer engineering, leveraging elements with dissimilar electron affinities, and enabling the integration of multiple functionalities and on-demand programmable magneto-ionics. Moreover, expanding the range of mobile species to include B broadens the scope of magneto-ionic applications, particularly in fields where boron plays a crucial role, such as electronics[26] or nuclear therapies[27] (*e.g.*, boron neutron capture therapy).

**Methods**

**Sample preparation**

50 nm-thick FeBO films with varying oxygen content were grown by DC magnetron sputtering using a Fe$_{80}$B$_{20}$ (nominal atomic %) target on Si(100) wafers, previously coated with 15 nm Ti, 50 nm Au, and 10 nm Ta. All depositions were carried out at room temperature using an AJA International ATC 2000-V Sputtering System with a base pressure < 1×10$^{-7}$ Torr. The oxygen content was changed by varying the O$_2$/Ar flow rate (*i.e.*, $\frac{O_2\ flow}{O_2\ flow\ +\ Ar\ flow}$) in the sputtering chamber: 0 (no oxygen), 2, and 5%; denoted as FeB, FeBO (2% O$_2$), and FeBO (5% O$_2$), respectively. As a standard procedure, the target was pre-sputter cleaned at a power of 100 W, whereas the sputter deposition of the films was performed at 55 W. The total pressure is fixed at 3×10$^{-3}$ Torr for the growth. The substrate-to-target distance was approximately 10 cm. Prior to the growth of the Ta buffer layer and FeB(O) films, the Au layer was partially masked to serve as the working electrode for subsequent magneto-electric characterization.

**Magneto-electric characterization**

Magnetic and magneto-electric measurements were carried out at room temperature using a vibrating sample magnetometer from MicroSense (LOT, Quantum Design). The magnetic fields were applied along the film plane direction. For standard hysteresis loops, the maximum applied field was 20 kOe. Magneto-electric measurements were carried out while applying voltages, using an external B2902A power source, between the counter electrode (a Pt wire) and the Au working electrode in a custom-made electrolytic cell[5,8], filled with liquid electrolyte—anhydrous propylene carbonate solvating Na$^+$ and OH$^-$ species (10 - 25 ppm). The magnetization was determined by normalizing the magnetic moment to the sample volume of the as-grown film exposed to the electrolyte. Any diamagnetic and paramagnetic contributions from the hysteresis loops were subtracted by correcting the linear slopes at high fields.

**Structural and compositional characterizations**

High-resolution transmission electron microscopy (TEM), high-angle annular dark-field scanning transmission electron microscopy (HAADF-STEM), and electron energy loss spectroscopy (EELS) were carried out on a Spectra 300 (S)TEM microscope (Thermo Fisher Scientific), operated at 200 kV. Prior to TEM observations, cross-sectional lamellae were prepared by focused ion beam, placed onto a copper transmission electron microscopy grid, and sputter-coated with a protective platinum layer.

Time of flight-energy (TOF-E) elastic recoil detection analysis (ERD)[28] was carried out to determine the Fe, B, and O concentrations along depth. An impinging ion beam of 12.6 MeV $^{79}$Br$^{6+}$ particles was used. The sample was tilted to a grazing incidence angle of 15° between the ion beam and the surface of the film. The elastic collision between the impinging nuclei and target atoms can result in recoiling target species. The mass, energy and probability of the atoms emitted from the sample in forward direction were recorded by means of a multi-dispersive detector telescope. ERD analysis yields a quantitative elemental depth profile of the sample. This technique probes atomic areal densities; hence the results are reported in units of $1\times10^{15}$ atom/cm$^2$. The depth profile is derived from the energy loss of the ions in matter.

X-ray absorption spectroscopy (XAS) at the B $K$ and Fe $L_{2,3}$ edges was carried out at the BL29-BOREAS beamline located at the ALBA Synchrotron[29], under ultra-high vacuum conditions (base pressure ~ $10^{-10}$ Torr). The spectra were collected at room temperature in fluorescence yield (FY) detection mode, with the incoming X-ray beam direction forming an angle of 15 degrees relative to the sample normal vector. The Fe $L_{2,3}$ edge XAS spectra were obtained from averaging 8 consecutive scans, whereas the B $K$ edge XAS data represent averages of 20 scans, to ensure quality datasets and to minimize any possible beam-related spectral shifts.

Defect characterization was carried out by variable energy positron annihilation lifetime spectroscopy (VEPALS). VEPALS measurements were conducted at the Mono-energetic Positron Source (MePS) beamline at Helmholtz-Zentrum Dresden – Rossendorf (Germany)[30]. A CeBr$_3$ scintillator detector together with a Hamamatsu R13089-100 photomultiplier tube for the gamma photons detection was employed. A Teledyne SPDevices ADQ14DC-2X digitizer with a 14-bit vertical resolution and 2GS/s (gigasamples per second) horizontal resolution was utilized for the processing of signals. The overall time resolution of the measurement system is ≈ 0.250 ns and all spectra contain at least $1\times10^7$ counts. A typical lifetime spectrum $N(t)$, which is the absolute value of the time derivative of the positron decay spectrum, is described by $N(t) = R(t) * \sum_{i=1}^{k+1} \frac{I_i}{\tau_i} e^{-t/\tau_i}$ + Background, where k is the number of different defect types contributing to the positron trapping, which are related to k + 1 components in the spectra with individual lifetimes $\tau_i$ and intensities $I_i$ ($\sum I_i = 1$). The instrument resolution function $R(t)$ is a sum of two Gaussian functions with distinct intensities and relative shifts both depending on the positron implantation energy, $E_P$. It was determined by measuring a reference sample, $i.e.$ yttria-stabilized zirconia, which exhibits a known single lifetime component of 182 ± 3 ps. The background was negligible. All the spectra were deconvoluted using a non-linear least-squares fitting method, minimized by the Levenberg-Marquardt algorithm in the software package PALSfit, into 2 major lifetime components, which directly evidence localized annihilation at 2 different defect types (sizes; $\tau_1$ and $\tau_2$). The shortest lifetime component $\tau_1$ represents smaller vacancy-clusters, while the lifetime component $\tau_2$ accounts for larger vacancy-clusters linked to grain boundaries or small pores. The relative intensity ($I_i$) of each component can be regarded to some extent as the concentration of each defect type[5,8]. In general, positron lifetimes may be overestimated in the topmost regions of films due to surface roughness and broken symmetry, but these effects are reduced beyond depths of 20 nm (> 2 keV in terms of positron implantation energy $E_P$, see below)[5]. Therefore, at depths exceeding 20 nm (> 2 keV), positron lifetimes can be considered fully representative of the bulk film properties. The positron lifetime and its intensity have been probed as a function of positron implantation energy $E_P$, which is related to the mean implantation depth $\langle z \rangle$ following: $\langle z \rangle \text{(nm)} = \frac{23.9}{\rho\left(\frac{g}{cm^3}\right)} E_P(\text{keV})^{1.69}$. $\langle z \rangle$ is an approximate measurement of depth since it does not account for positron diffusion.

**Electrical characterization**

To assess the electrical properties, 50-nm FeB(O) films were deposited directly onto high-resistive $SiO_2$/Si substrates, following the same sputtering condition used for depositing magneto-ionic heterostructures. The electrical transport measurements were carried out at room temperature using the van der Pauw configuration. For as-grown FeB, FeBO (2% $O_2$) and FeBO (5% $O_2$) films, resistivity values of $2.0\times10^{-4}$, $2.7\times10^{-4}$, and 1.7 ohm cm are obtained, respectively.

**Additional information**

Correspondence and requests for materials should be addressed to E.M.

**Author contributions**

Z.M. and E.M. had the original idea and designed the experiments. Z.M., J.S., and E.M. supervised the work. E.M. led the investigation. Z.M., A.Q., and E.P. synthesized the heterostructures. Z.M., J.N., and J.S. carried out the magneto-electric measurements and analyzed the data. K.-A.K. and J.M. conducted the ERD measurements and analyzed the data. Z.M. and J.H.-M. carried out the XAS characterization and analyzed the data. Z.M., A.Q., J.N., and J.S. performed the TEM and STEM characterization and carried out the analysis of the corresponding data. M.O.L., E.H., and A.W. characterized the samples by PALS and analyzed the data. Z.M., H.T., and E. M. performed the transport property measurements and analyzed the data. All authors discussed the results and commented on the article. The article was written by Z.M. and E.M. with contributions from all co-authors.

**Competing financial interests**

The authors declare no competing financial interests.

**Acknowledgements**

Financial support by the European Union's Horizon 2020 Research and Innovation Programme (BeMAGIC European Training Network, ETN/ITN Marie Skłodowska–Curie grant N° 861145), the European Research Council (2021-ERC-Advanced REMINDS Grant N° 101054687), the Spanish Government (PID2020-116844RB-C21, TED2021-130453B-C21, and TED2021-130453B-C22), the

Generalitat de Catalunya (2021-SGR-00651), and the MCIN/AEI/10.13039/501100011033 & European Union NextGenerationEU/PRTR (grant Nº CNS2022-135230) is acknowledged. ICN2 is funded by the CERCA program/Generalitat de Catalunya. The ICN2 is supported by the Severo Ochoa Centres of Excellence program, Grant CEX2021-001214-S, grant funded by MCIU/AEI/10.13039/501100011033. Authors acknowledge the use of instrumentation financed through Grant IU16-014206 (METCAM-FIB) to ICN2 funded by the European Union through the European Regional Development Fund (ERDF), with the support of the Ministry of Research and Universities, Generalitat de Catalunya. The XAS experiments were performed at BL29-BOREAS beamline at ALBA Synchrotron with the collaboration of ALBA staff under the proposal with ID 2024028386. VEPALS were carried out at ELBE from the Helmholtz-Zentrum Dresden – Rossendorf e. V., a member of the Helmholtz Association. We would like to thank the facility staff for their assistance. E. M. is a Serra Húnter Fellow.

# Supplementary Information

**Table of contents**

**Supplementary Fig. S1: The influence of oxygen on the microstructure of FeBO films.**

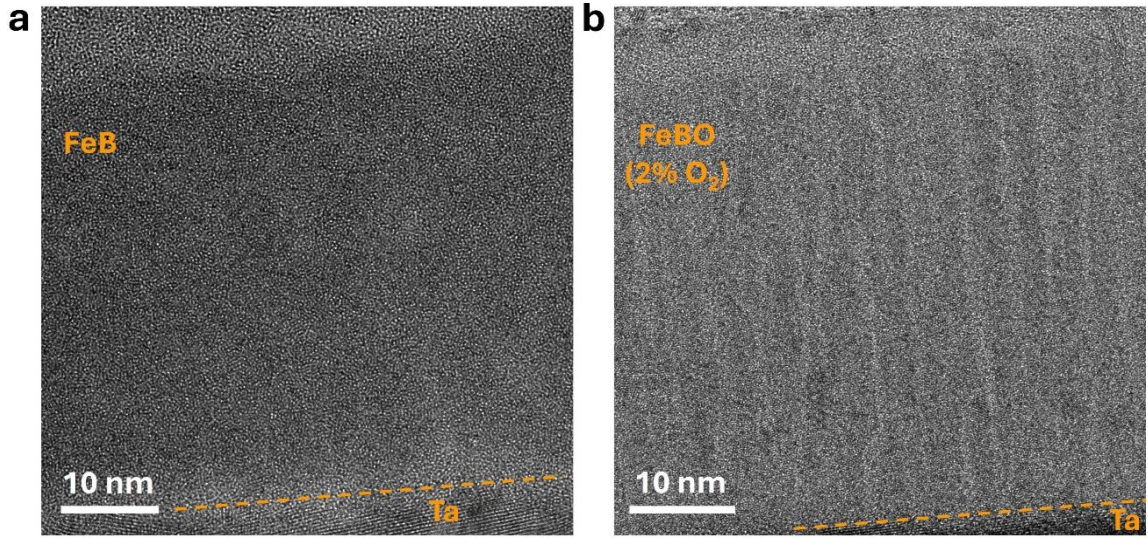

**Fig. S1** | Representative TEM micrographs of the cross-sections of as-grown **a**, FeB, and **b**, FeBO (2% $O_2$) films, both are 50 nm in thickness, showing the feature of columnar growth when the heterostructures are sputtered in an oxidized atmosphere.

**Supplementary Fig. S2: Further TEM characterization of as-grown FeBO (2% O$_2$) films.**

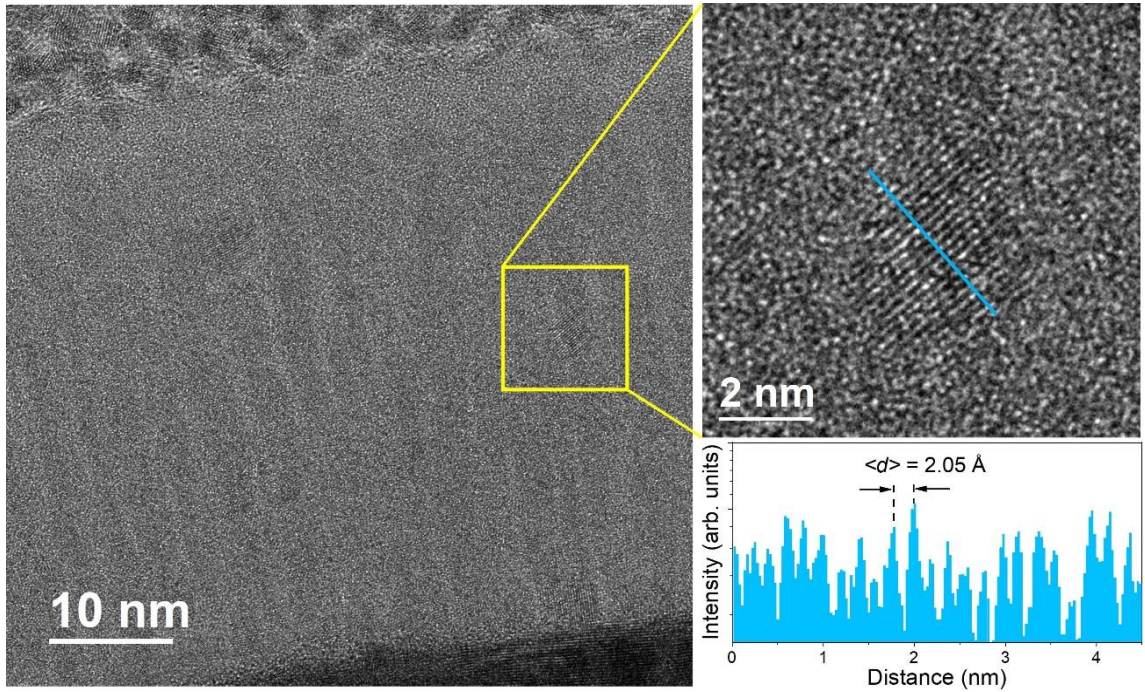

**Fig. S2** | Further TEM characterization of as-grown FeBO (2% O$_2$) films. Crystallites compatible with Fe can be observed in certain regions as shown in the enlarged view of the marked yellow square (PDF® 00-001-1262).

**Supplementary Fig. S3: Magnetic hysteresis loops of the as-grown and gated FeB and FeBO (2% O₂) films.**

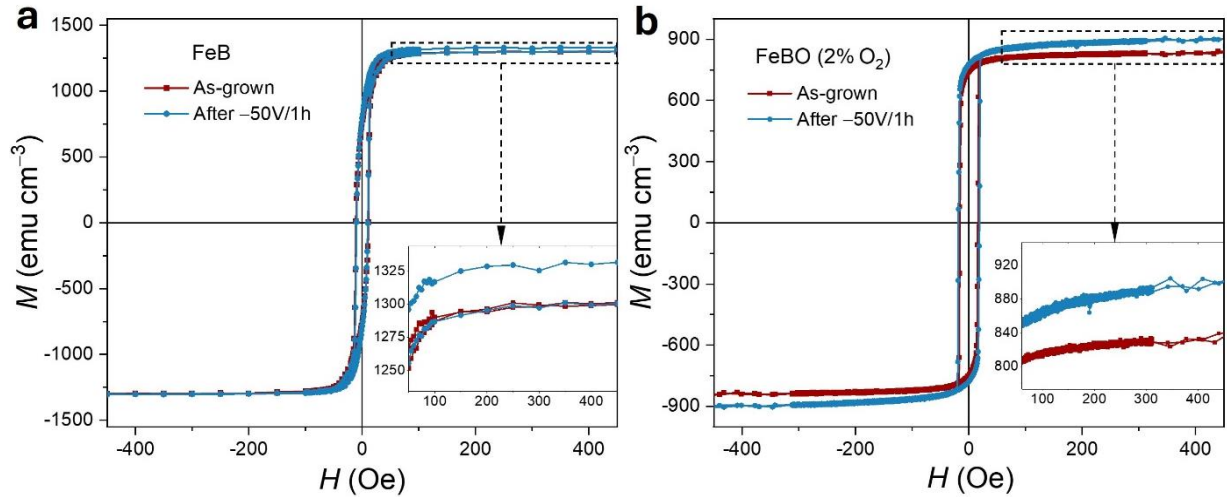

**Fig. S3 | a**, Hysteresis loops of the as-grown and gated (–50 V for 1 h) FeB films. **b**, Hysteresis loops of the as-grown and gated (–50 V for 1 h) FeBO (2% O₂) films. In these loops, the magnetization evolution at high fields up to 450 Oe is shown. The volatility of the voltage-induced magnetization increase is evident from the inset of **a** (*i.e.*, after gating, the 1st branch of the loop exhibits a $M_S$ of 1330 emu cm$^{-3}$, which drops to about 1300 emu cm$^{-3}$ in the 4th branch of the loop, the same value as that for the as-grown sample). In contrast, for the FeBO (2% O₂) film, the magnetization enhancement remains non-volatile (*i.e.*, no sign of $M_S$ decrease is observed from the loop recorded after gating).

**Supplementary Fig. S4: Magnetic hysteresis loops of the as-grown and gated FeO (2% O₂) films.**

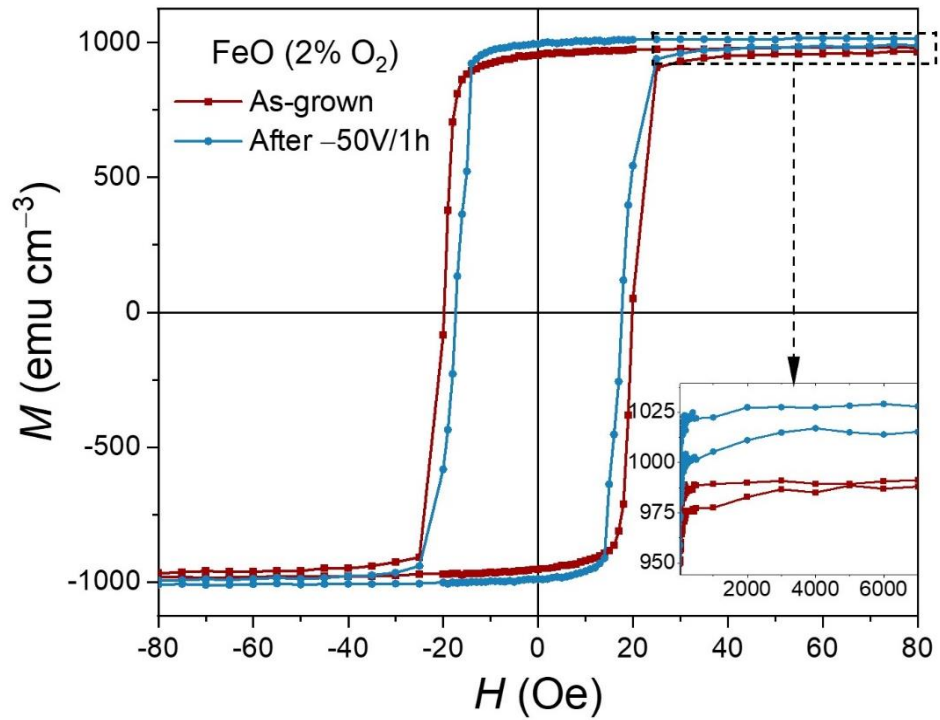

**Fig. S4 |** Hysteresis loops of the as-grown and gated (–50 V for 1 h) FeO (2% O₂) films. The inset presents a zoom-in of the high-field region marked by the dashed rectangle. Note the slight change (≈ 2%) change in $M_S$ in the voltage-treated sample in the return branch of the loop.

**Supplementary Fig. S5: Depth-dependence of the relative atomic percentage of B with respect to the B and Fe atomic percentages of the as-grown and gated FeB and FeBO (2% O2) films.**

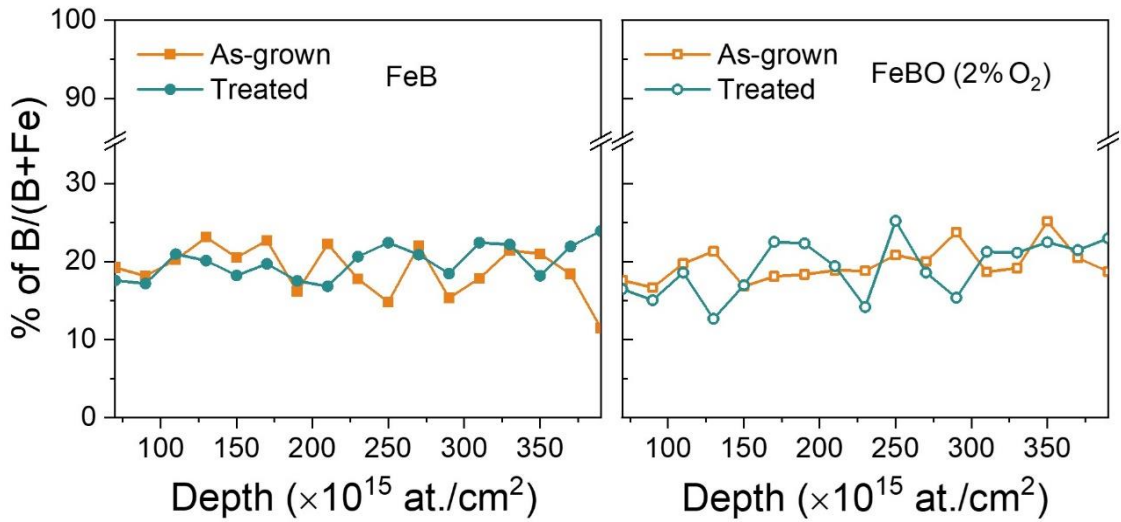

**Fig. S5** | Evolution of the atomic percentage of B (% of B/(B+Fe)) as a function of depth for FeB and FeBO (2% $O_2$) films, shown in the left and right panels, respectively. Note that the B/Fe ratio does not significantly change in depth, in agreement with a dual-cation movement of Fe and B ions during the electrolyte gating process.

**Supplementary Fig. S6: Magnetic hysteresis loops of the as-grown and gated FeBO (5% O$_2$) films.**

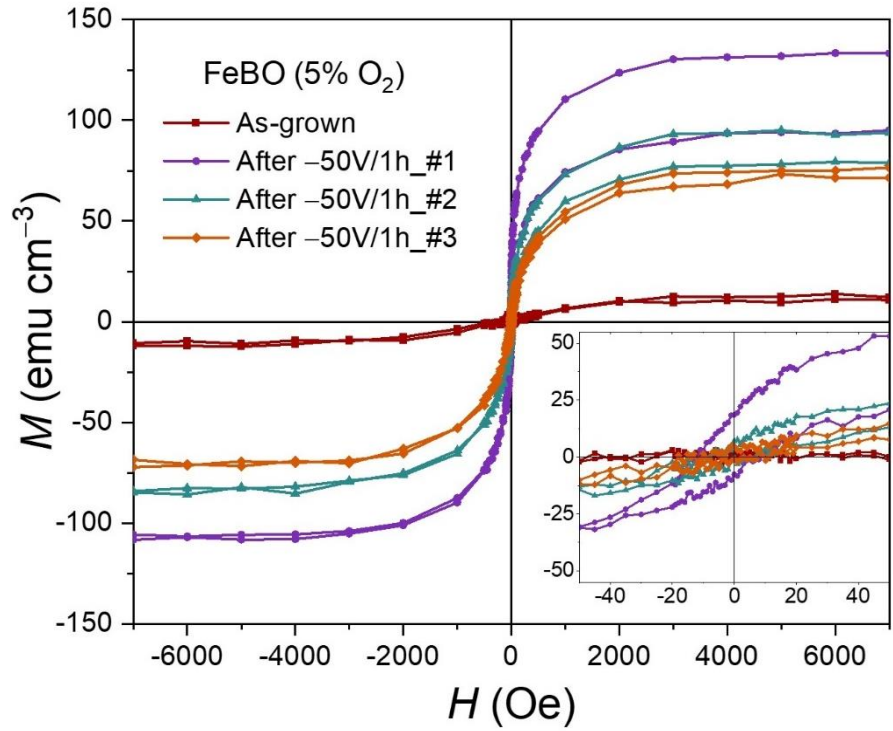

**Fig. S6** | Hysteresis loops of the as-grown and gated (–50 V for 1 h) FeBO (5% O$_2$) films. Upon gating, consecutive hysteresis loops were recorded until a permanent ferromagnetic response is reached.

**Supplementary Fig. S7: FY-XAS spectra of as-grown FeB, FeBO (2% $O_2$), and FeBO (5% $O_2$) films.**

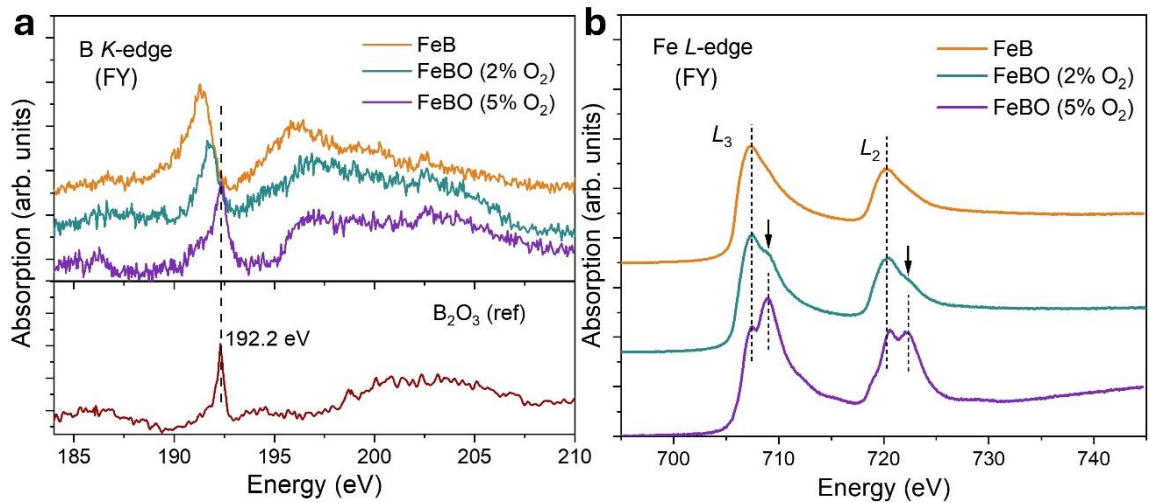

**Fig. S7** | Fluorescence yield (FY) XAS spectra at the **a**, B *K*-edge, and **b**, Fe *L*-edge for the as-grown FeB, FeBO (2% $O_2$), and FeBO (5% $O_2$) films. The bottom panel of **a** presents the $B_2O_3$ reference spectrum (peaked at 192.2 eV), collected from a pyrex borosilicate glass. The gradual shift of the B *K*-edge peak indicates the increased B oxidation. The occurrence of the shoulder peaks at higher energy ends in FeBO (2% $O_2$) implies a slight increase in Fe oxidation. In the FeBO (5% $O_2$) film, the shoulder peaks become stronger, indicating much enhanced oxidation state.

**Supplementary Fig. S8: Fe and B depth profiles by TOF-E ERD of the as-grown and gated FeBO (5% $O_2$) films.**

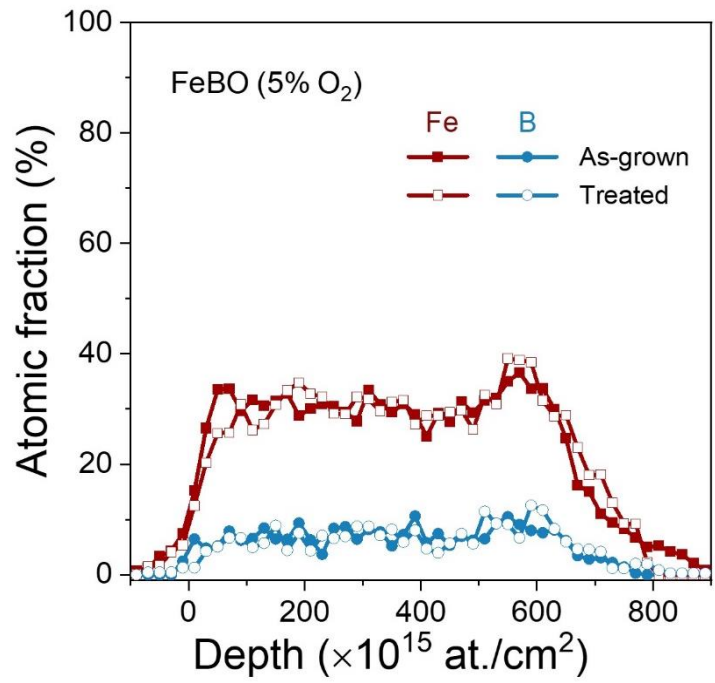

**Fig. S8** | Fe and B depth profiles by TOF-E ERD for the as-grown and gated FeBO (5% $O_2$) films.

**Supplementary Note 1: Correlation between positron lifetime and the number of Fe vacancies ($V_{Fe}$).**

For as-grown and voltage-treated FeB and FeBO (2% $O_2$) heterostructures, $\tau_1$ values correspond to vacancy-clusters ranging from 3 to 8 Fe vacancies ($V_{Fe}$ in Fig. 4). With respect to $\tau_2$, which represents larger vacancy-clusters, clusters larger than 16 Fe vacancies ($V_{Fe}$ in Fig. 4) are observed in both as-grown and voltage-treated FeB and FeBO (2% $O_2$) samples[1]. To use Fe as an analogue here is justified as evidenced by the TEM FFT analysis indicating that the phase is close to metallic Fe (Fig. 1c).

**Supplementary References**